\documentclass[sigconf, screen, nonacm]{acmart} % nonacm: CC BY 4.0 block, no 'Not for redistribution'

\newif\ifanonymous
\anonymousfalse

\ifanonymous
  \newcommand{\ToolEval}{[EvalTool]}
  \newcommand{\ToolGen}{[GenTool]}
  
  \newcommand{\selfcite}[1]{[anonymous]}
  \newcommand{\selfciteTwo}[2]{[anonymous]}
  \newcommand{\selfciteThree}[3]{[anonymous]}
  
\else
  \newcommand{\ToolEval}{JudgeGPT}
  \newcommand{\ToolGen}{RogueGPT}
  
  \newcommand{\selfcite}[1]{\cite{#1}}
  \newcommand{\selfciteTwo}[2]{\cite{#1, #2}}
  \newcommand{\selfciteThree}[3]{\cite{#1, #2, #3}}
  
\fi

\usepackage{fancyhdr}

\AtBeginDocument{%
  }

\copyrightyear{2026}
\acmYear{2026}
\setcopyright{cc}
\setcctype{by}
\acmDOI{10.1145/3795513.3807431}

\acmConference[WebSci Companion '26]{18th ACM Web Science Conference}{May 26--29, 2026}{Braunschweig, Germany}
\acmBooktitle{18th ACM Web Science Conference (WebSci Companion '26), May 26--29, 2026, Braunschweig, Germany}
\acmISBN{979-8-4007-2492-3/2026/05}

\begin{document}

\title{Can Humans Tell? A Dual-Axis Study of Human Perception of LLM-Generated News}
\titlenote{Published in the Companion Publication of the 2026 18th ACM Web Science Conference (WebSci Companion '26), pp.~34--39. DOI: \href{https://doi.org/10.1145/3795513.3807431}{10.1145/3795513.3807431}.}

%%
%% Author information
%%
\author{Alexander Loth}
\authornote{Corresponding author.}
\email{alexander.loth@stud.fra-uas.de}
\orcid{0009-0003-9327-6865}
\affiliation{%
  \institution{Frankfurt University of Applied Sciences}
  \city{Frankfurt am Main}
  \country{Germany}
}

\author{Martin Kappes}
\email{kappes@fra-uas.de}
\orcid{0000-0002-8768-8359}
\affiliation{%
  \institution{Frankfurt University of Applied Sciences}
  \city{Frankfurt am Main}
  \country{Germany}
}

\author{Marc-Oliver Pahl}
\email{marc-oliver.pahl@imt-atlantique.fr}
\orcid{0000-0001-5241-3809}
\affiliation{%
  \institution{IMT Atlantique, UMR IRISA, Chaire Cyber CNI}
  \city{Rennes}
  \country{France}
}

\renewcommand{\shortauthors}{Loth, et al.}

\begin{abstract}
Can humans tell whether a news article was written by a person or a large language model (LLM)? We investigate this question using \ToolEval{}, a study platform that independently measures source attribution (human vs.\ machine) and authenticity judgment (legitimate vs.\ fake) on continuous scales. From 2,318 judgments collected from 1,054 participants across content generated by six LLMs, we report five findings: (1)~participants cannot reliably distinguish machine-generated from human-written text ($p > .05$, Welch's $t$-test); (2)~this inability holds across all tested models, including open-weight models with as few as 7B parameters; (3)~self-reported domain expertise predicts judgment accuracy ($r = .35$, $p < .001$) whereas political orientation does not ($r = -.10$, n.s.); (4)~clustering reveals distinct response strategies (``Skeptics'' vs.\ ``Believers''); and (5)~accuracy degrades after approximately 30 sequential evaluations due to cognitive fatigue. The answer, in short, is no: humans cannot reliably tell. These results indicate that user-side detection is not a viable defense and motivate system-level countermeasures such as cryptographic content provenance.
\end{abstract}




\maketitle

%%% ARXIV PREPRINT HEADER
\pagestyle{fancy}
\fancyhf{}
\fancyhead[L]{\textit{A.\ Loth, M.\ Kappes, M.-O.\ Pahl}}
\fancyhead[R]{\textit{Published in WebSci Companion '26, pp.\ 34--39}}
\fancyfoot[C]{\thepage}
\renewcommand{\headrulewidth}{0pt}

\section{Introduction}
\label{sec:introduction}

Text has long been the web's default trust carrier, yet LLMs now produce synthetic text that humans rate as no less credible than human-written content~\cite{kreps2022all, clark2021all}. As false and misleading content already spreads faster than corrections online~\cite{vosoughi2018spread}, the ability to generate convincing fabrications at marginal cost amplifies what Wardle and Derakhshan term ``information disorder''~\cite{wardle2017information} across an already fragile ecosystem~\cite{ferrara2024genai, lazer2018science, tandoc2018defining}.

Automated detection approaches face a moving-target problem: classifiers trained on one model generation degrade rapidly on the next~\cite{chen2023combating, su2024adapting}. A complementary line of research therefore focuses on the human side, investigating cognitive inoculation~\cite{kozyreva2023critical, roozenbeek2022psychological} and digital literacy interventions~\cite{lewandowsky2021countering}. Designing such interventions, however, requires empirical data on \emph{how} and \emph{where} human judgment fails.

Building on an earlier survey of the research landscape~\selfcite{loth2024blessing}, we developed \ToolEval{}~\selfcite{loth2026eroding}, a web-based study platform with two methodological innovations. First, it decouples \emph{authenticity} (legitimate vs.\ fake) from \emph{source} (human vs.\ machine) via independent continuous scales, avoiding the conflation inherent in binary real/fake tasks. Second, stimuli are generated by \ToolGen{}~\selfciteTwo{loth2026collateraleffects}{loth2026eroding}, a controlled multi-model framework that enables systematic comparison across LLM families. This poster reports five empirical findings from 2,318 judgments and discusses their implications for web system design.

\section{Related Work}
\label{sec:related}

Research on AI-generated misinformation spans two complementary directions. On the detection side, neural classifiers such as Grover~\cite{zellers2019defending} and more recent transformer-based detectors~\cite{chen2023combating} initially achieved high accuracy but degrade as models evolve~\cite{ferrara2024genai}; comprehensive surveys confirm this trend~\cite{zhou2020survey}. On the human side, controlled experiments consistently show that participants perform near chance when distinguishing machine-generated from human-written text~\cite{jakesch2023heuristics, clark2021all}. Kreps et al.\ demonstrated that GPT-generated political messages are rated as credible as journalist-written ones~\cite{kreps2022all}, while Su et al.\ showed that detection difficulty increases with model scale~\cite{su2024adapting}.

Most human-subject studies, however, employ binary classification tasks (real vs.\ fake) that conflate source and veracity: a machine-generated paraphrase of a true story is not misinformation, yet binary framing treats it as such. Our dual-axis method addresses this limitation. We additionally build on work in psychological inoculation~\cite{roozenbeek2022psychological, kozyreva2023critical} and analytic thinking as a predictor of misinformation resilience~\cite{pennycook2021psychology}, while connecting to emerging provenance standards (C2PA) as system-level alternatives~\selfcite{loth2026originlens}.

\section{Methodology}
\label{sec:methodology}

\ToolEval{} is a web-based study platform for collecting human perception data on AI-generated text~\selfcite{loth2026eroding}. Its design was informed by a systematic gap analysis of prior work~\selfcite{loth2024blessing}.

\begin{figure}[h!]
    \centering
    \includegraphics[width=0.95\columnwidth]{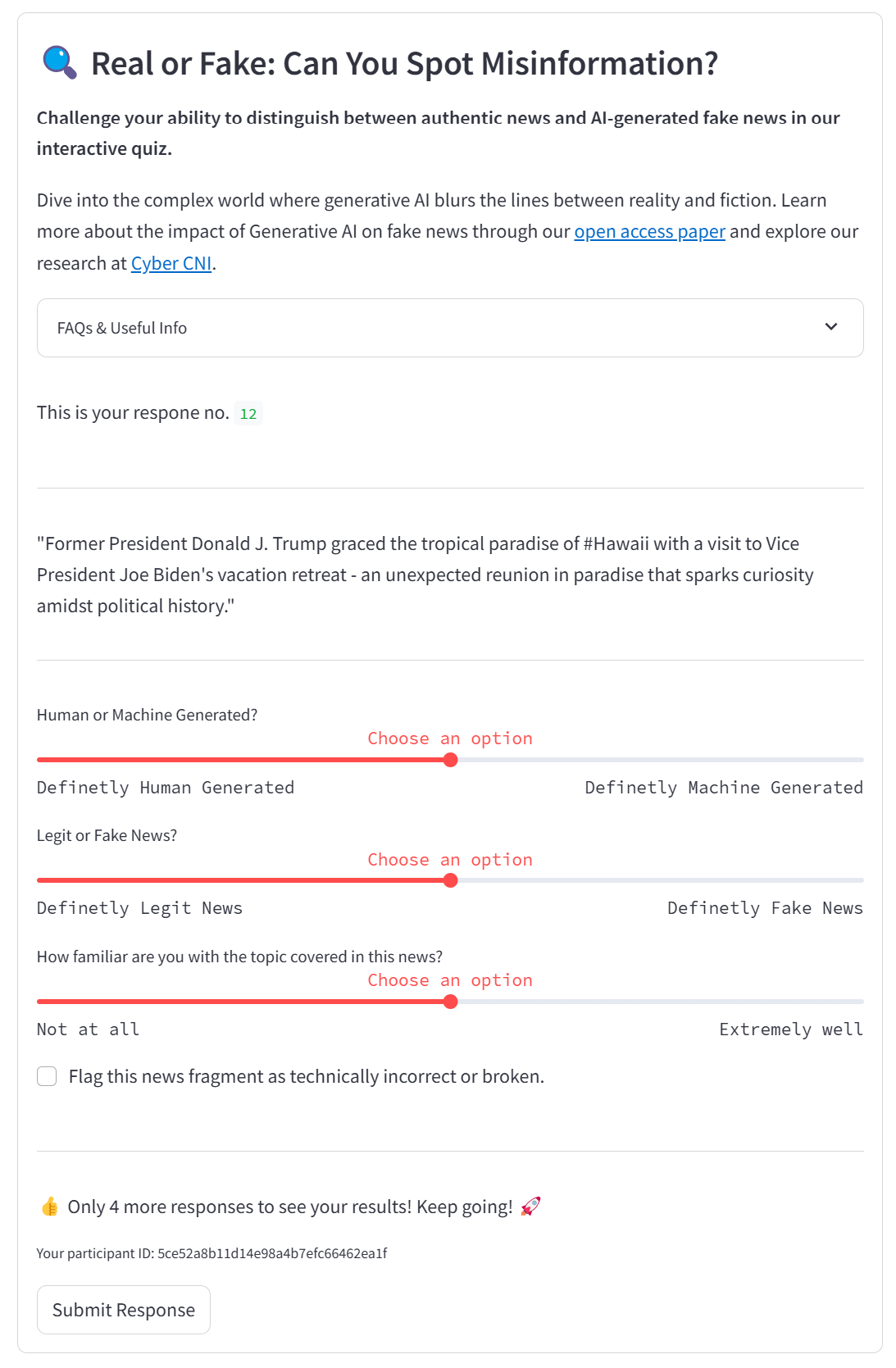}
    \caption{The \ToolEval{} interface. Participants rate each fragment on three continuous axes: source attribution, authenticity, and topic familiarity.}
    \label{fig:judgegpt-ui}
\end{figure}

\paragraph{Dual-Axis Assessment.}
For each news fragment, participants provide three independent ratings on continuous 0--100 sliders (Figure~\ref{fig:judgegpt-ui}): (1)~\emph{Source judgment} (0 = certainly machine, 100 = certainly human), (2)~\emph{Authenticity judgment} (0 = certainly fake, 100 = certainly legitimate), and (3)~\emph{Topic familiarity}. Continuous scales were chosen over Likert items to capture degree of certainty and to enable parametric analyses including Pearson correlation and clustering.

\paragraph{Stimulus Design.}
Stimuli follow a $2 \times 2$ factorial design crossing \emph{origin} (human, machine) with \emph{veracity} (real, fake). Machine-generated fragments are produced by \ToolGen{}~\selfciteTwo{loth2026collateraleffects}{loth2026eroding}, which orchestrates six LLMs (GPT-4~\cite{bubeck2023sparks}, GPT-3.5, GPT-4o, LLaMA-2~13B, Gemma~7B, Mistral~7B) with persona-based prompting. Human-sourced fragments are drawn from established news outlets and known misinformation databases. AI-generated texts are grounded in real news topics and verified for factual consistency to avoid conflating perception of style with detection of hallucinations. The stimulus set is intentionally skewed toward machine-origin fragments (${\sim}$98\%), with human-origin items serving as calibration anchors. This design choice reflects the study's focus on \emph{within-AI} variation (across models) rather than \emph{human-vs.-AI} comparison; participants are not informed of the base rate, and the near-chance detection results (Finding~1) hold when analyzed on the human-origin subset alone.

\paragraph{Data Collection.}
Participants provide informed consent and complete a demographic questionnaire (age, education, political orientation, AI familiarity) before evaluating a sequence of fragments. Each participant evaluates between 5 and 87 fragments (median = 12, IQR = 6--22), with presentation order randomized and model assignment balanced across participants. The platform records the three slider ratings, response latency, and an anonymous participant identifier for linking judgments to covariates. The sample skews toward educated European demographics (68\% university-educated, 74\% European); we discuss generalizability limitations in Section~\ref{sec:discussion}.

\section{Findings}
\label{sec:findings}

The dataset comprises 2,318 judgments from 1,054 unique participants. Table~\ref{tab:summary} provides an overview; each finding is detailed below.

\begin{table}[h!]
\centering
\caption{Summary of empirical findings ($N = 2{,}318$ judgments from $1{,}054$ participants; median 12 items per participant, range 5--87).}
\label{tab:summary}
\small
\begin{tabular}{p{0.22\columnwidth}p{0.34\columnwidth}p{0.32\columnwidth}}
\toprule
\textbf{Finding} & \textbf{Statistic} & \textbf{Implication} \\
\midrule
F1: Undetectable & $t(2316) = 0.87$, $p = .38$ & Intuition fails as defense \\
F2: Model-agnostic & All 6 LLMs at $\bar{x} \approx 0.50$ & Low barrier to deceptive text \\
F3: Expertise $>$ Politics & $r = .35$ vs.\ $r = -.10$ & Literacy-based interventions \\
F4: User personas & $k = 2$ clusters (sil. $= .41$) & Adaptive designs needed \\
F5: Fatigue & Decline after ${\sim}$30 items & Sustained vigilance infeasible \\
\bottomrule
\end{tabular}
\end{table}

\begin{figure}[h!]
    \centering
    \includegraphics[width=0.95\columnwidth]{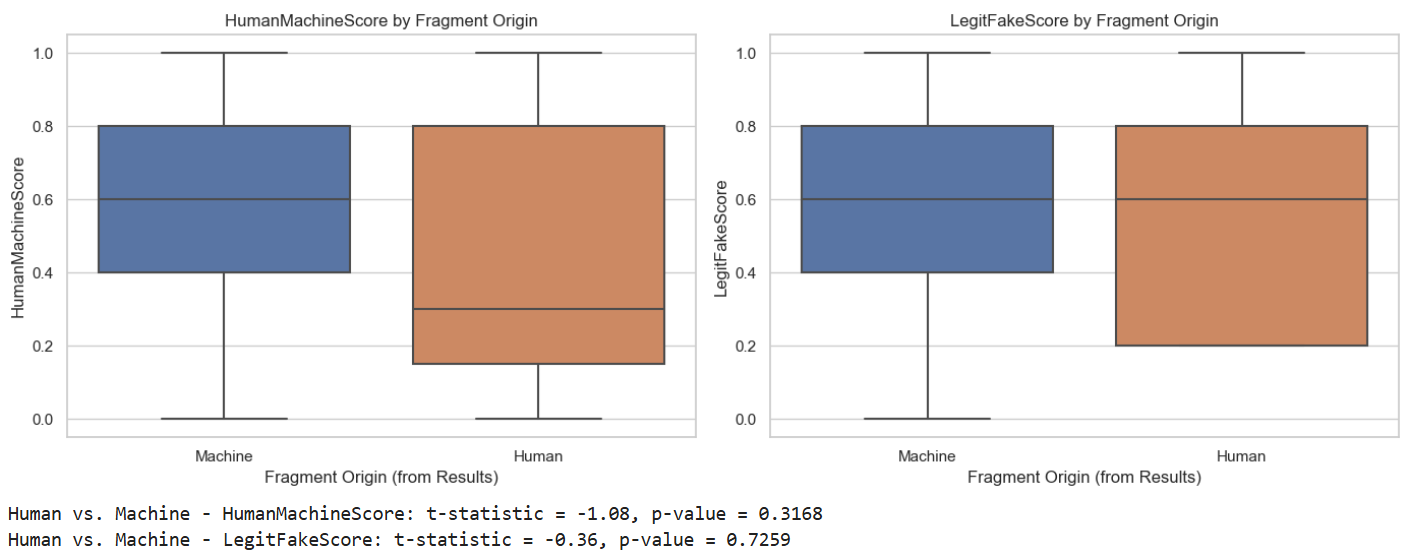}
    \caption{Source score (left) and authenticity score (right) distributions for machine- vs.\ human-written fragments. Overlapping distributions and a non-significant $t$-test indicate participants cannot distinguish the two conditions.}
    \label{fig:human_vs_machine}
\end{figure}

\paragraph{Finding 1: Machine-Generated Text Is Not Detectable by Participants.}
Figure~\ref{fig:human_vs_machine} shows near-complete overlap between human- and machine-generated content on both axes. A Welch's $t$-test yields no significant difference in source scores between conditions ($t(2316) = 0.87$, $p = .38$, Cohen's $d = 0.04$), confirming that participants' intuitions do not reliably discriminate AI-generated from human-written text.

\begin{figure}[h!]
    \centering
    \includegraphics[width=0.95\columnwidth]{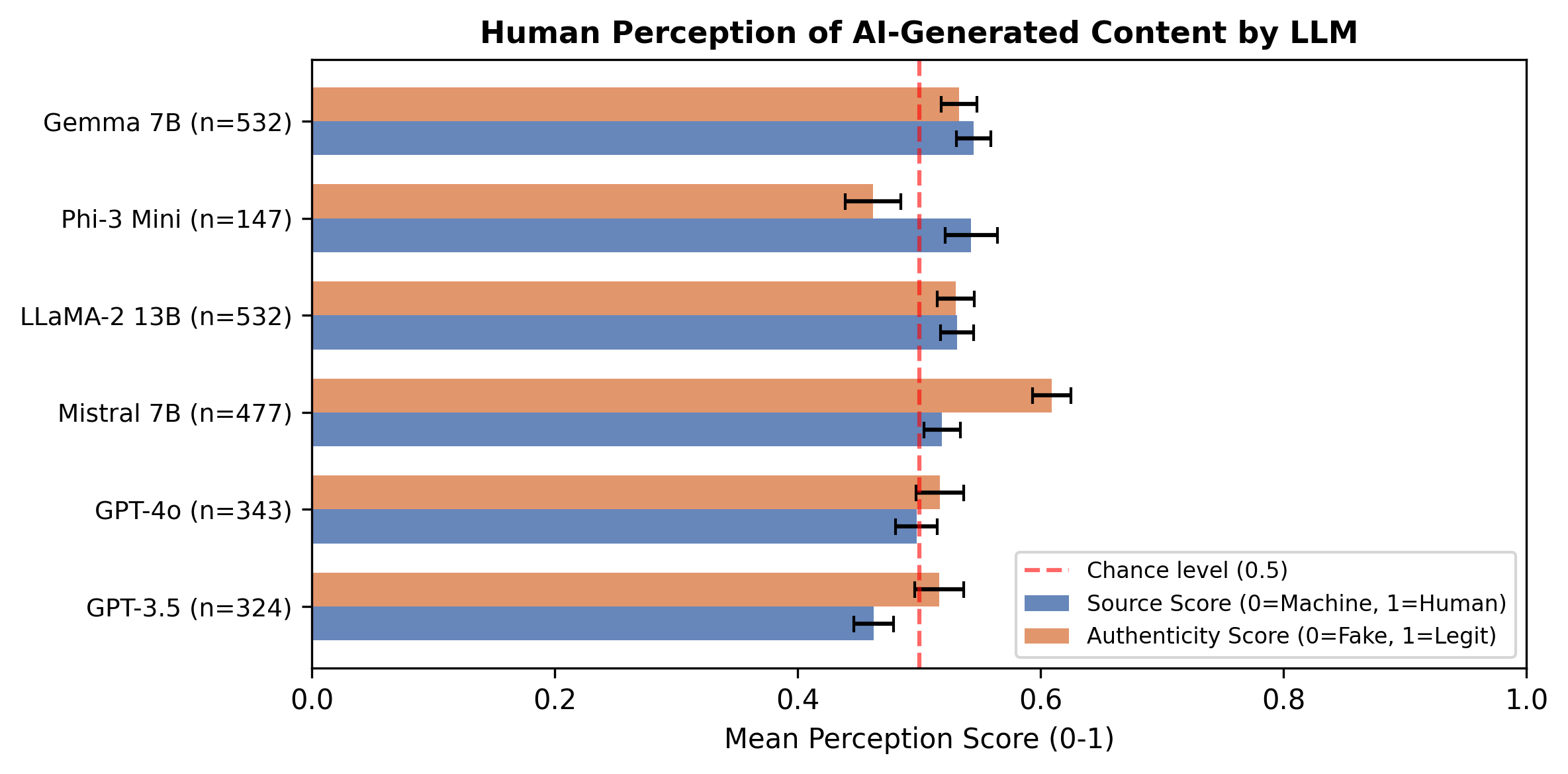}
    \caption{Mean source and authenticity scores per LLM ($\pm$ SE). The dashed line marks chance level (0.5). No model is reliably identified as machine-generated.}
    \label{fig:model_perception}
\end{figure}

\paragraph{Finding 2: Detection Failure Is Model-Agnostic.}
Figure~\ref{fig:model_perception} disaggregates source scores by generating model. Mean scores for all six LLMs fall within the 0.44--0.55 range, clustering around the 0.5 chance level. A one-way ANOVA reveals no significant between-model effect ($F(5, 2349) = 1.92$, $p = .09$). Open-weight models with as few as 7B parameters produce text rated no differently from GPT-4o output, indicating that the capability to generate human-indistinguishable text is no longer restricted to frontier models.

\begin{figure*}[t]
    \centering
    \includegraphics[width=0.95\textwidth]{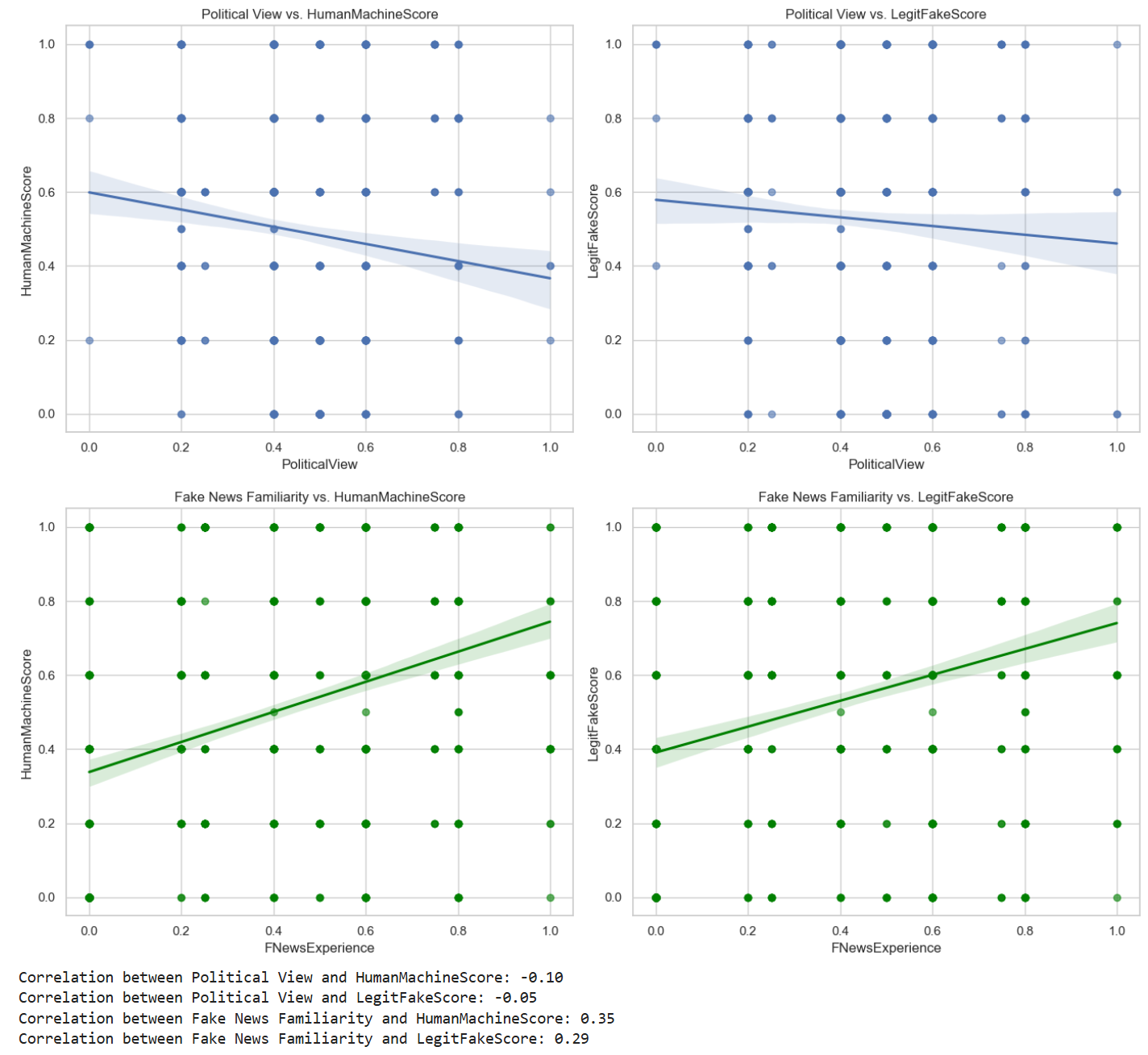}
    \caption{Relationships between participant covariates and judgment scores. Top: political view (weak slope). Bottom: fake news familiarity (positive slope). Regression lines with 95\% CI shown.}
    \label{fig:correlations}
\end{figure*}

\paragraph{Finding 3: Domain Expertise Predicts Accuracy; Political Orientation Does Not.}
Pearson correlations (Figure~\ref{fig:correlations}) show that self-reported fake news familiarity is positively associated with source judgment accuracy ($r = .35$, $p < .001$) and authenticity accuracy ($r = .29$, $p < .001$). Political extremity shows only a weak, non-significant negative correlation ($r = -.10$, $p = .12$). While the political orientation measure is coarse (a single self-report item) and the stimulus set was not designed to control for politically sensitive content, the pattern suggests that learned analytical skills may be a stronger predictor of detection performance than ideological predisposition. A dedicated study with politically balanced stimuli would be needed to draw causal conclusions about the role of political orientation.

\begin{figure*}[t]
    \centering
    \includegraphics[width=0.95\textwidth]{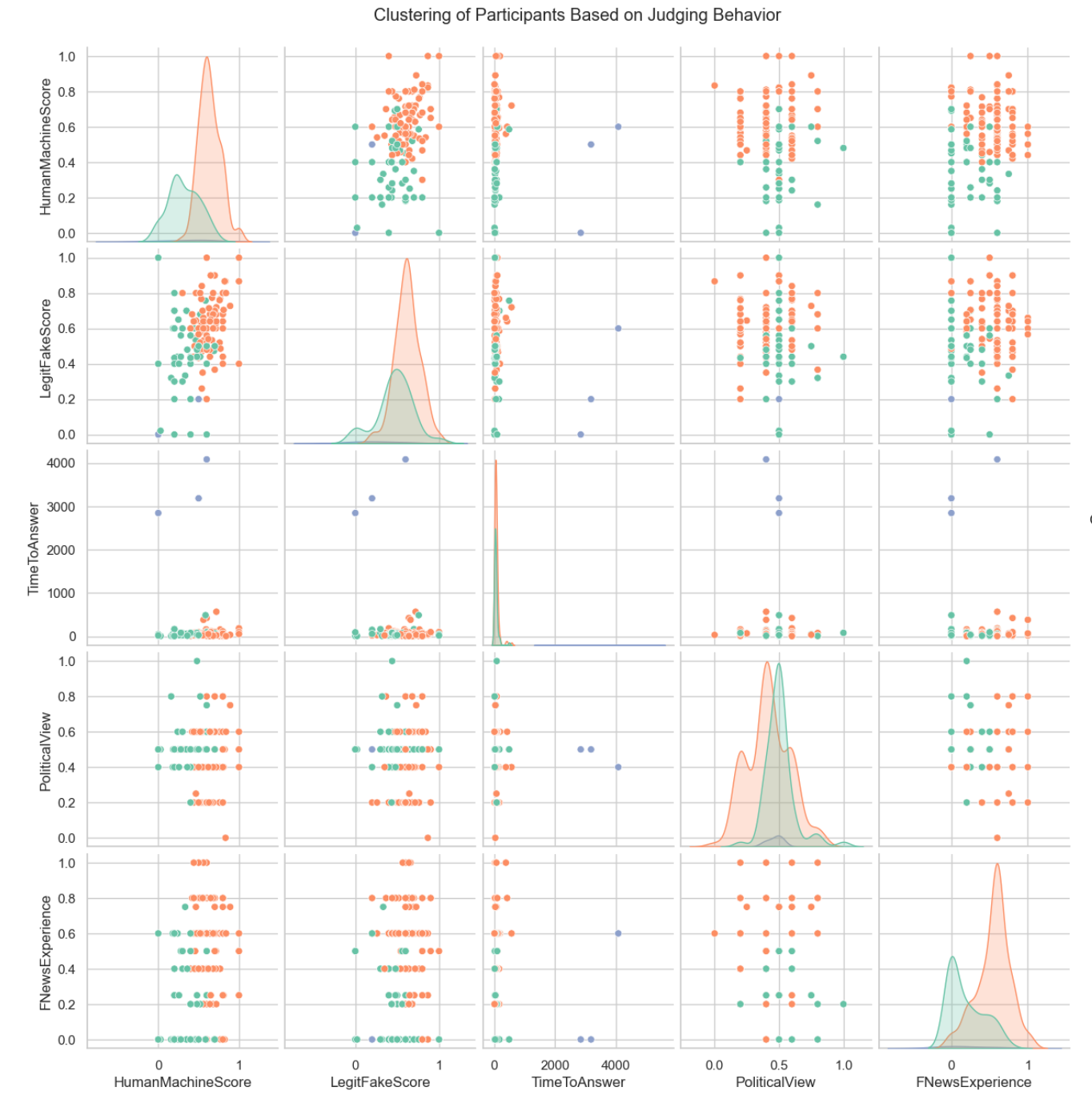}
    \caption{Pairplot of participant-level mean scores colored by cluster. The two groups show clearly separated distributions on both axes.}
    \label{fig:clusters}
\end{figure*}

\paragraph{Finding 4: Clustering Reveals Distinct Response Strategies.}
$K$-means clustering ($k = 2$, selected via silhouette analysis, silhouette coefficient $= .41$) on participant-level mean scores identifies two groups (Figure~\ref{fig:clusters}): ``Skeptics,'' who assign low trust across all content regardless of origin, and ``Believers,'' who maintain high baseline trust. These distinct response strategies imply that uniform interventions will be suboptimal.

\begin{figure*}[t]
    \centering
    \includegraphics[width=0.95\textwidth]{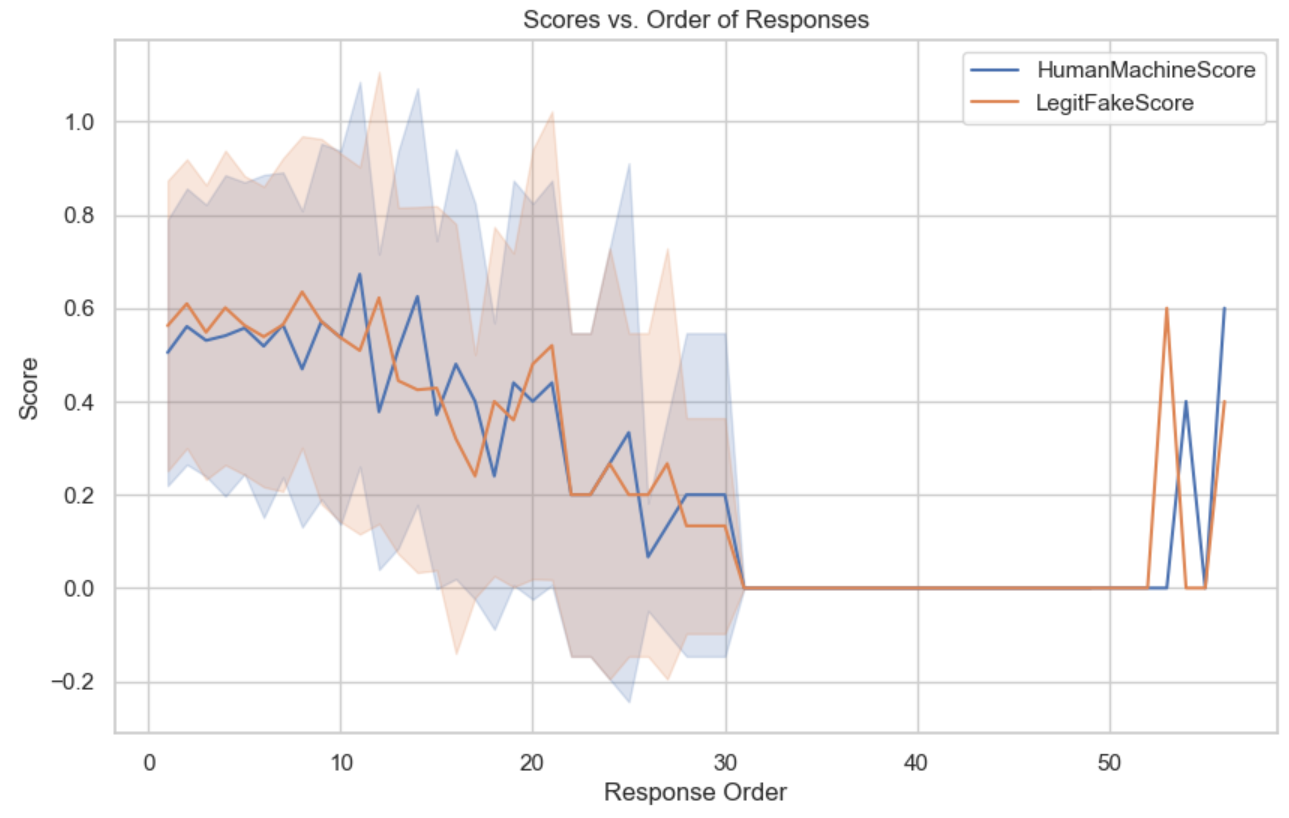}
    \caption{Rolling mean of judgment scores across sequential evaluations. Scores improve initially before declining after ${\sim}$30 items, consistent with a learning-then-fatigue pattern.}
    \label{fig:fatigue}
\end{figure*}

\paragraph{Finding 5: Initial Learning Effect Followed by Cognitive Fatigue.}
A rolling-window analysis of sequential judgments (Figure~\ref{fig:fatigue}) reveals improved accuracy during the first 15--20 evaluations, consistent with a learning effect. Beyond approximately 30 items, accuracy declines and participants increasingly default to ``fake'' classifications, indicating cognitive fatigue. This temporal pattern limits the practical duration of detection-based interventions.

\section{Discussion and Implications}
\label{sec:discussion}

Three design implications follow from our findings:

\begin{itemize}
    \item \textbf{Provenance over detection.} The failure of human judgment across all tested models (F1, F2) implies that detection cannot be offloaded to users. Cryptographic provenance frameworks such as C2PA, which attach verifiable content histories at the point of creation, offer a more scalable alternative~\selfcite{loth2026originlens, loth2026originlenswebsci}, complemented by source-level credibility signals~\selfcite{loth2026cred1}.

    \item \textbf{Bounded inoculation.} The learning effect (F5, first 15--20 items) supports the efficacy of inoculation-based interventions~\cite{roozenbeek2022psychological, kozyreva2023critical}, but the subsequent fatigue effect constrains their practical deployment to short, focused sessions rather than continuous monitoring.

    \item \textbf{Persona-aware design.} The Skeptic/Believer distinction (F4) suggests that web platforms should adapt trust indicators to user disposition rather than applying uniform labels. The expertise effect (F3) further implies that interventions building analytical skill may be more productive than those targeting partisan bias, though this finding should be validated with politically balanced stimuli in future work.
\end{itemize}

These findings are part of a broader empirical program on AI-driven disinformation encompassing generation infrastructure~\selfcite{loth2026collateraleffects}, human perception~\selfcite{loth2026eroding}, expert assessment~\selfcite{loth2026verification}, provenance-based countermeasures~\selfcite{loth2026originlens, loth2026originlenswebsci}, domain credibility infrastructure~\selfcite{loth2026cred1}, and a unified dissertation framework formalizing the ``Indistinguishability Threshold''~\selfcite{loth2026symposium}.

\paragraph{Limitations.} The participant sample skews toward educated European demographics (68\% university-educated, 74\% European), limiting generalizability to broader populations. The controlled evaluation format does not capture the social and contextual cues present in real-world social media encounters. The stimulus set is intentionally skewed toward machine-generated content (${\sim}$98\%); while this design serves the study's focus on within-AI variation across models, a balanced design would strengthen claims about human-origin anchoring effects and control for potential base-rate adaptation by participants. The political orientation finding (F3) is preliminary: the single-item measure and lack of politically controlled stimuli preclude causal claims. Future work will address these limitations through a longitudinal browser-extension study with a more diverse participant pool and balanced stimulus design.

\section{Conclusion}
\label{sec:conclusion}

Can humans tell? Our dual-axis study of 2,318 judgments across six LLM families provides a clear empirical answer: they cannot. Machine-generated text is indistinguishable from human writing regardless of model size or family, domain expertise predicts detection accuracy more strongly than political orientation, participants adopt distinct trust strategies, and cognitive fatigue limits sustained detection. These findings support a shift from user-level detection toward system-level countermeasures, including content provenance, adaptive trust indicators, and bounded inoculation interventions.

\paragraph{Data Availability.} All datasets are archived on Zenodo under restricted academic access: RogueGPT stimulus corpus (DOI: \href{https://zenodo.org/records/18703138}{10.5281/zenodo.18703138}), JudgeGPT human perception data (DOI: \href{https://zenodo.org/records/18703385}{10.5281/zenodo.18703385}), and the CRED-1 domain credibility dataset (DOI: \href{https://zenodo.org/records/19355308}{10.5281/zenodo.19355308}). The \ToolEval{} platform\footnote{\url{https://github.com/aloth/JudgeGPT}} and \ToolGen{} framework\footnote{\url{https://github.com/aloth/RogueGPT}} are open-source.

\begin{acks}
We thank all participants of the \ToolEval{} study.
\end{acks}

\bibliographystyle{ACM-Reference-Format}
\bibliography{bibliography-arxiv-v2}

\end{document}